\documentclass[12pt]{iopart}

\usepackage{cite}
\usepackage{graphicx}
\usepackage{graphics}
\usepackage{iopams}
\usepackage{color}
\usepackage{doi}

\begin{document}

\title{Critical behavior of the 2D Ising model modulated by the Octonacci sequence}
\author{G.A. Alves$^{1}$, M.S. Vasconcelos$^{2}$ and T.F.A. Alves$^{3}$}
\address{$^{1}$Departamento de F\'{i}sica, Universidade Estadual do Piau\'{i}, 59078-900, Teresina - PI, Brazil}
\address{$^{2}$Escola de Ci\^encias e Tecnologia, Universidade Federal do Rio Grande do Norte, 59078-900, Natal - RN, Brazil}
\address{$^{3}$Departamento de F\'{\i}sica, Universidade Federal do Piau\'{i}, 57072-970, Teresina - PI, Brazil}
\eads{alves.gladstone@gmail.com, manoel\_vasconcelos@hotmail.com, tay@ufpi.edu.br}

\begin{abstract}

We investigated the Ising model on a square lattice with ferro and
antiferromagnetic interactions modulated by the quasiperiodic
Octonacci sequence in both directions of the lattice. We have
applied the Replica Exchange Monte Carlo (Parallel Tempering)
technique to calculate the thermodynamic quantities of the system.
We obtained the order parameter, the associated magnetic
susceptibility ($\chi$) and the specific heat $(c)$ in order to
characterize the universality class of the phase transition. Also,
we use the finite size scaling method to obtain the critical
temperature of the system and the critical exponents $\beta$,
$\gamma$ and $\nu$. In the low temperature limit we have obtained a
continuous transition with critical temperature around $T_{c}
\approx 1.413$. The system obeys the Ising universality class with
logarithmic corrections. We found estimatives for the correction
exponents $\hat{\beta}$, $\hat{\gamma}$ and $\hat{\lambda}$ by using
the finite size scaling technique.

\end{abstract}


\submitto{Journal of Statistical Mechanics: Theory and Experiment (JSTAT)}

\maketitle

\section{Introduction}

The discovery of an Al-Mn quasicrystal by  Shechtman et al.\cite{PhysRevLett.53.1951}, awarded with the Nobel Prize, and the pioneering work of Merlin et al.\cite{PhysRevLett.55.1768} on the non periodic Fibonacci and Thue-Morse GaAs-AlAs superlattices, have created a new and promising research field, namely the {\it physics of quasicrystals}. They are a particular type of solid that have a non Bravais discrete point-group symmetry, like a C5 symmetry in two dimensions, or icosahedral symmetry in three dimensions. A more precise definition of quasicrystals with dimensionality $d$ ($d=$ 1, 2 or 3) has been given recently \cite{NatPhoton.7.177}. In addition to their possible generation by a substitution process, they can also be formed from a projection of an appropriate periodic structure into a higher dimensional space ($D=m$), where $D$ is the dimensionality and $m>d$.

In the last four decades, many high-quality Al-based quasicrystals have been developed\cite{JPhysColloques.47.1986,JapanJApplPhys.26.1987,MTJapanInstMet.30.1989,JMaterRes.6.1991} and characteristic physical properties of these quasicrystals have been investigated\cite{JPSJ.62.639,JNonCrystSolids.153.325,JNonCrystSolids.153154.573}. Among their properties, it is well known that the magnetic properties of alloys containing a magnetic element are very sensitive to their local atomic structures such as the kind of the nearest-neighbor atoms, the atomic distance and the coordination number. Indeed, the magnetic properties of quasicrystals are very different from those of amorphous alloys in spite of the similarity properties, exhibited by them, of both local and atomic structures\cite{MatsubaraZNA.43.1988}. In the last three decades, the study of quasicrystals have significantly advanced the knowledge about the atomic scale structure\cite{Steurer:sc5003,C3CS35388E}. However, many questions about magnetic properties, regarding the consequences of quasiperiodicity on physical properties, remain open. For example, an unanswered question is if the long-range antiferromagnetic (AFM) order can be sustained in real quasicrystalline systems.

Due to the fact that quasicrystals have no translational symmetry, many anomalous properties different from those of a regular crystal are expected. In fact, quasicrystals are essentially different from disordered materials. The main difference between them and the disordered materials is the fact that they have self-similar properties, i.e., any finite section of the quasicrystals is reproduced within a distance of a degree of its linear (2D or 3D) scale. Also, it is well known that this system exhibit long-range correlations \cite{PhysRevB.57.2826}. Therefore, at sufficiently low temperatures, where long-range correlation is expected to play an important role, one may expect that some physical quantities will ``reflect'' the quasiperiodicity of the lattice. In fact, in contrary to the well known spin structure in antiferromagnets in periodic lattices, the antiferromagnetic arrangement in quasicrystals has been in constant scientific debate in the past ten years (for a review see \cite{PhiMaga.86.733} and references therein). It has been shown that rare earth-containing quasicrystals exhibit an aperiodic diluted ferrimagnet freezing phase at low temperatures \cite{PhysRevB.57.R11047,SCHEFFER2000629}. These studies suggest that this new magnetic state is an intermediate between a canonical spin glass and a typical superparamagnet \cite{JPhysCondMatt.46.7981}.

On the other hand, a substantial number of theoretical works have considered the possibility of non-trivial ordering of localized magnetic moments on quasicrystals\cite{PhysRevLett.92.047202,PhysRevLett.93.076407,Matsuo2004421,PhysRevB.71.115101,PhilosMag.86.733}. These studies have concluded, generally, that the answer is affirmative. However, to date, no quasi-antiferromagnets have been discovered. But, it is known that the behavior of some quasicrystals in low-temperature have magnetic topological order and frustration\cite{MatsubaraZNA.43.1988}, as in spin glass behavior\cite{Hattori0953}. This fact allow us the opportunity to study new (simple) theoretical models to answer the open questions or give some insight on this subject. For the critical behavior, the change of the critical exponents is partially answered by the Harris-Luck criterion, valid for ferromagnetic systems, described by exchange terms of the form $J(1+\Delta)$ where $\Delta \ll 1$ is modulated by a quasiperiodic letter sequence\cite{EurophysLett.24.359}.

In a general way, by growth processes, it is possible to control the disorder in a given system and move progressively from a long-range structural order to a quenched disorder \cite{PhysRevB.61.15738,RevModPhys.71.1125} or, alternatively, to a quasiperiodic order by modifying the exchange strengths and signals. Specifically, the quasiperiodic order in quasicrystals can be realized in three ways: 1) we can change the interactions or 2) the geometry of the crystal lattice 3) or both. Here we follow 2) by considering a square lattice with ferro and antiferromagnetic exchange interactions modulated by a quasiperiodic sequence.

Recently, a quasiperiodic model based on Fibonacci sequence with quasiperiodic long-ranged order with competing interactions have been published by us\cite{PhysRevE.93.042111}. We have obtained the critical behavior of a second order transition, with critical temperature $T_{c}\approx 1.268$ \cite{PhysRevE.94.019904}. An interesting characteristic of the disordered antiferromagnetic system is the frustration\cite{PhysRevE.89.042139}. It plays a role in those systems by inducing an aperiodic diluted ferrimagnet phase. It is known, for the 2d square lattice, if the system is underfrustrated (when frustrated plaquettes are removed in a controlled way), the disordered system presents spin glass phase at low temperatures\cite{PhysRevE.89.042139}, otherwise, for the pure stochastic frustration we have only a paramagnetic phase. This opens the opportunity to find other quasiperiodic systems that could exhibit unusual orderings at low temperatures. Therefore, one of the objectives of this work is to obtain the thermodynamic properties of the Ising model in two dimensions with positive and negative exchange interactions with the same strength, ordered by a quasiperiodic Octonacci sequence in both directions of a square lattice.

The Octonacci sequence can be built from the Ammann-Beenker tiling, which is an octagonal tiling obtained by using a strip projection method \cite{crystallography.126.2009} (in fact, Ammann bars for the Ammann-Beenker tiling have distances between them according an Octonacci sequence). The name Octonacci comes from \textit{``Octo''} for orthogonal and \textit{``nacci''} from the Fibonacci sequence, the oldest example of a quasiperiodic chain. Here we call attention to the fact that this sequence can be confused with the first Fibonacci generalization sequence called {\it silver mean}. The unique (and great) difference between those sequences is that Octonacci sequence is symmetric with respect to its substitution rule $A\rightarrow ABA$, while the {\it silver mean} is not symmetric, since it follows the rule $A\rightarrow AAB$ \cite{Aperiodic}. Apparently, it is not a great difference, but it will affect the quasiperiodicity and consequentially, the long-range correlations can be different, mainly at high generations of the sequence.

This paper is organized as follows: In section 2 we describe the model and the Hamiltonian. The order parameter $q$, magnetic susceptibility $\chi$, specific heat $c$ and the critical behavior of the system are shown in section 3 and finally we present some conclusions and general comments in section 4.

\section{Model and Simulations}

We consider the Ising model in a square lattice with only first neighbor interactions, where the Hamiltonian is given
by\cite{ZPhys.31.253}
\begin{equation}
\mathcal{H}=-\sum_{\langle i,j \rangle}J_{ij}S_{i}S_{j},
\end{equation}
where $S_{i}$ and $S_{j}$ are the spins on sites  $i$ and $j$, respectively and their values can be $\pm 1$. $J_{ij}$ is the exchange interaction strength between first neighbor spins $S_{i}$ and $S_{j}$. The exchange constants at the lattice bonds can be modulated according to an aperiodic letter sequence. When $J_{ij}$ is $1$, we have a ferromagnetic interaction, and when $J_{ij}$ is $-1$ we have an antiferromagnetic one.

The Octonacci sequence is obtained from the substitution rules
$A\rightarrow ABA$ and $B\rightarrow A$ in 1D
\cite{OptMater.46.378,OptMater.62.584,OptMater.64.126}. Any
generation of the aperiodic sequence can be constructed from the
previous generation by replacing all letters \textit{A} with
\textit{ABA} and all letters \textit{B} with \textit{A}. Starting
with the letter \textit{A}, by repetitive applications of the
substitution rule we can obtain the successive iterations of the
Octonacci sequence in 1D.

\begin{figure}[h!]
\begin{center}
\includegraphics[scale=0.38]{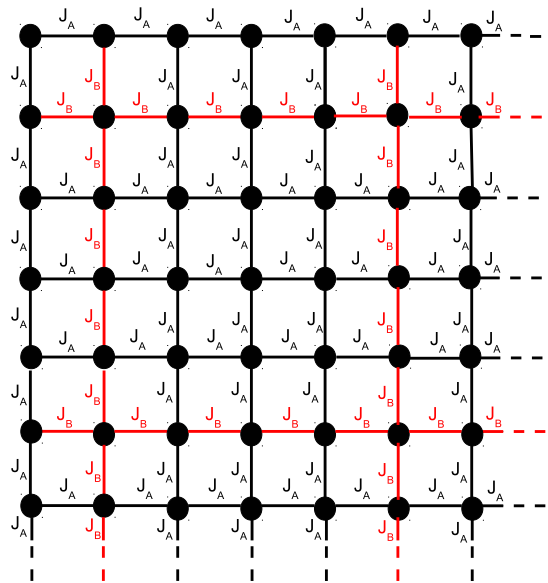}
\end{center}
\caption{Example of a lattice with exchange interactions modulated by the Octonacci sequence. The black and red lines stand for exchange interaction strengths $J_{A}=1$ (ferromagnetic) and $J_{B}=-1$ (antiferromagnetic) respectively. We used the Octonacci letter sequence, which is obtained from the substitution rules $A\rightarrow ABA$ and $B\rightarrow A$ which means that the any generation of the lattice can be constructed from the previous generation by replacing all letters \textit{A} with \textit{ABA} and all letters \textit{B} with \textit{A}. The horizontal bonds follow a Octonacci sequence in the vertical direction and similarly for the vertical bonds in order to produce frustration.}
\label{fig_quasiperiodic_lattice}
\end{figure}

To modulate the exchange interactions in a square lattice, we set
the exchange strengths in the horizontal lattice bonds by the
Octonacci sequence at the vertical direction and conversely for the
horizontal direction. In this way we can obtain frustrated lattice
plaquettes and we can expect a change of the critical behavior at
lower temperatures. We show an example of such lattice in
Fig.(\ref{fig_quasiperiodic_lattice}).

By using the Replica Exchange Monte Carlo technique (also known as
Parallel Tempering)\cite{PhysChemChemPhys.7.3910,
PhysRevLett.57.2607,JPSJ.65.1604,Geyer.1991}, which is suited to
approach the problem of determining the steady state of systems with
complex energy landscapes composed of many local minima and to find
the ground state of systems with non-periodic interactions, we
obtained the staggered magnetization order parameter $\langle q
\rangle$, the associated susceptibility $\chi$, the specific heat
$c$ and Binder cumulant $g$
\begin{eqnarray}
q & = & \frac{1}{N^*}\sum_{i}^{N} S^{0}_{i}S_{i} \label{s0s_parameter}\\
\chi & = & N\left( \langle q^{2} \rangle - \langle q \rangle^{2} \right)/T, \label{susceptibility}\\
c & = & N\left(\langle \mathcal{H}^{2} \rangle - \langle \mathcal{H} \rangle^{2}\right)/T^{2}, \label{specific_heat} \\
g & = & 1-\frac{\langle q^{4} \rangle}{3\langle q^{2} \rangle^{2}}, \label{binder_cumulant}
\end{eqnarray}
where $\langle...\rangle$ stands for a thermal average over sufficiently many independent steady state system configurations, $S^{0}_{i,j}$ is the ground state of the system and $L$ and $T$ are the lattice size and the absolute temperature, respectively. We used the following values of the lattice size $L$: 17, 41, 99 and 239, which are Pell's numbers $P_n$, given by the recursion rule
\begin{equation}
P_n = 2P_{n-1} + P_{n-2},
\label{pellrecursionrule}
\end{equation}
where $P_{0}=1$ and $P_{1}=1$. The total number of spins for each lattice size is $N=L^{2}$. In the calculation of the order parameter, we set $N^*$ as the size of the ordered cluster at $T \rightarrow 0$ to ensure a proper behavior of the order parameter, which is the limit $q \rightarrow 1$ in the $T \rightarrow 0$ limit\cite{PhysicaA.363.327}. In fact, the frustrated plaquettes are a fraction of the total plaquettes, and that fraction depends only on the substitution rule. So, the number of undecided spins, responsible for the degeneracy of the steady state, which are excluded from the ordered cluster at $T \rightarrow 0$, and $N^*=N-N_u$ itself, are a fraction of $N$. We have tracked the number $N_u$ of undecided spins for each Octonacci generation, namely: $N_u = 8$, $120$, $696$ and $4060$ for $L=17$, $41$, $99$ and $239$ ($N=L^{2}$), respectively. So, we have $N^*=a_iN$, where $a_i \approx 0.972$, $0.929$, $0.929$ and $0.929$ for $N=17$, $41$, $99$ and $239$, respectively. We can see that, using $N^*$ in place of $N$, we have the same of multiplicating the definitions by a constant factor $1/a$, i. e.,
\begin{equation}
q = \frac{1}{aN}\sum_{i=1}^{N}S_{i}^{0}S_{i}
\end{equation}
with $N^{*}=aN$. This does not change any scaling properties, just to ensure $\lim_{T \rightarrow 0} q = 1$.

To determine the critical behavior, we have used the following
Finite Size Scaling (FSS) relations\cite{PhysA.391.1753}, with
logarithmic corrections\cite{PhysRevLett.96.115701,PhysRevLett.97.155702,PhysRevE.82.011145,kenna:2012}
\begin{eqnarray}
q & \propto & L^{\beta/\nu}\left(\ln{ L }\right)^{\hat{\beta}+\beta\hat{\lambda}}f_q(\vartheta), \label{q_fss} \\
\chi & \propto & L^{-\gamma/\nu}\left(\ln{ L }\right)^{\hat{\gamma}-\gamma\hat{\lambda}}f_{\chi}(\vartheta), \label{susceptibility_fss} \\
c & \propto & \left( \ln L \right)^{-\hat{\alpha}}f_{c}(\vartheta), \label{specificheat_fss} \\
g & \propto & f_{g}(\vartheta), \label{cumulant_fss}
\end{eqnarray}
where $\beta=1/8$, $\gamma=7/4$, $\alpha=0$ (logarithmic divergence)
and $\nu=1$ are the critical exponents (the Ising 2d ones). The
$\hat{\alpha}$, $\hat{\beta}$, $\hat{\gamma}$ and $\hat{\lambda}$
are the logarithmic correction exponents. The $f_{i}(\vartheta)$ are
the FSS functions with a logarithmic corrected scaling variable
\begin{equation}
\vartheta=L^{1/\nu}\left(T-T_{c}\right)\left| \ln \left| T-T_{c}\right|\right|^{-\hat{\lambda}}.
\end{equation}
The correction exponents $\hat{\alpha}$, $\hat{\beta}$, $\hat{\gamma}$ and $\hat{\lambda}$ obey the following scaling relations\cite{kenna:2012}
\begin{equation}
\hat{\alpha} = 1-d\nu\hat{\lambda}
\label{fss_scaling_relations1}
\end{equation}
\begin{equation}
2 \hat{\beta} - \hat{\gamma} = -d\nu\hat{\lambda},
\label{fss_scaling_relations2}
\end{equation}
where $d$ is the dimensionality of the system. The scaling relation
(\ref{fss_scaling_relations1}) is valid only for $\alpha = 0$
(logarithmic divergences), in the general case, $\hat{\alpha} =
-d\nu\hat{\lambda}$. For  $\alpha = 0$ and $\hat{\alpha} = 0$, we
have the double logarithmic divergence ($\ln \ln L$) of the specific
heat as seen for the 2d diluted Ising model\cite{kenna:2012}.

We used $1 \times 10^{6}$ MCM (Monte-Carlo Markov) steps to make the $N_t=400$ system replicas (each system replica has a different temperature) reach the equilibrium state and the independent steady state system configurations are estimated in the next $1 \times 10^{6}$ MCM steps with $10$ MCM steps between one system state and another one to avoid self-correlation effects. Every MCM steps are composed of two parts, a sweep and a swap. One sweep is accomplished when all N spins were investigated if they flip or not and one swap is accomplished if all the $N_t$ lattices are investigated if they exchange or not their temperatures (swap part). We carried out $10^{5}$ independent steady state configurations to calculate the
needed thermodynamic averages.

\section{Results and Discussion}

\begin{figure}[h!]
\begin{center}
\includegraphics[scale=0.8]{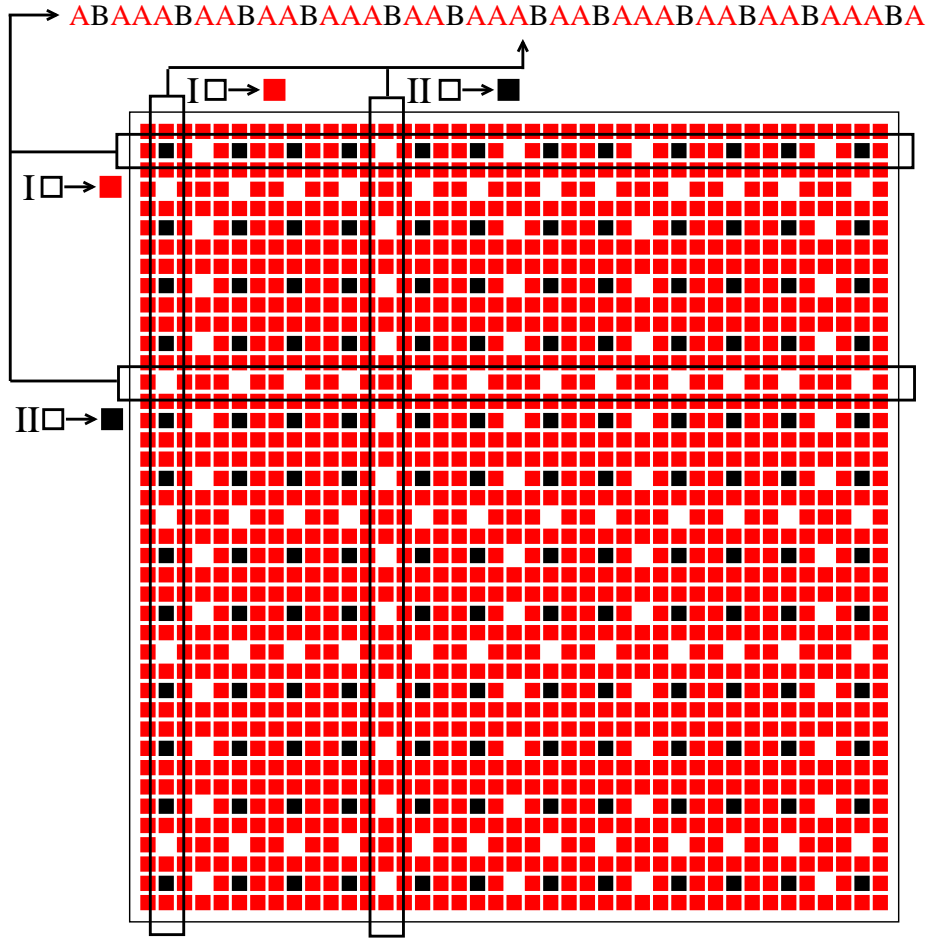}
\end{center}
\caption{(Color online) Ground state of the Hamiltonian for a system
size $L=41$. Grey (red) squares indicate a spin with ``up''
orientation and black squares indicate a spin with ``down''
orientation. Blank squares indicate an ``undecided'' spin, where its
relative orientation does not change the ground state energy. We see
that the Column/Line I type, when ``undecided'' spins take the
``up'' orientation, the column/line resembles the octonacci
sequence. The same is true for the Column/Line II type, if
``undecided'' spins take the  ``down'' orientation. The ground state
is non-periodic, but clearly follows the same ordering as indicated
in a way we can construct the ground states for all the lattice
sizes.} \label{fig_steadystate}
\end{figure}

First, we show the ground state of the model in
Fig.(\ref{fig_steadystate}) for a system size $L=41$. In the figure,
squares in grey (red in color version) indicate a spin with ``up''
orientation and squares in black indicate a spin with ``down''
orientation (and vice-versa). Squares in blank, indicate an
``undecided'' spin, where its relative orientation does not change
the ground energy (the ground state still preserves $Z_2$ symmetry).
We note the ground state is non-periodic (its follows Octonacci
sequence in a given direction) and highly degenerated due to the
frustrations present in the lattice. We can see that this peculiar
ordering is an aperiodic ferrite (a periodic arrangement of ``up''
and ``down'' spins) which, at first, could be described by a
staggered magnetization, namely
\begin{equation}
m^* = \frac{1}{N}\sum_{i=1}^{N} e^{i\vec{q}_i \cdot \vec{r}_i}S_i,
\end{equation}
where the spins are $S_i=\pm 1$ and the phase factors are $S^{0}_i =
e^{i\vec{q}_i \cdot \vec{r}_i} = \pm 1$. The application of this
order parameter to our case can be done if the ground state is given
by a substitution rule as we will justify by the following
reasoning: First, by noting that the staggered magnetization is a
magnetization calculated at sublattices, and the phase factors can
take values $S^{0}_{i}=\pm 1$, depending on the site in the lattice,
when measuring $m^*$ directly at the simulation, the value is zero
because $S_i=\pm 1$ values have the same probability. So, instead of
this, we should measure $m = \left| m^* \right|$ in the simulation.
To decouple the modulus in the lattice summation (here we must
express a modulus of a sum as a sum of the moduli), we should
decouple the sublattices $j$ where $S^0_i=S^0_j$ with $S^0_j=1$ or
$S^0_j=-1$ for every sublattice site $i$ and, then, express the
sublattice magnetization as $m_j = S^{0}_j\left| m^*_j \right|$.
Thus, the staggered magnetization is given by a sum of sublattice
magnetizations $m = \sum_{j=1}^{k} m_j$.

\begin{figure}[h!]
\begin{center}
\includegraphics[scale=0.5]{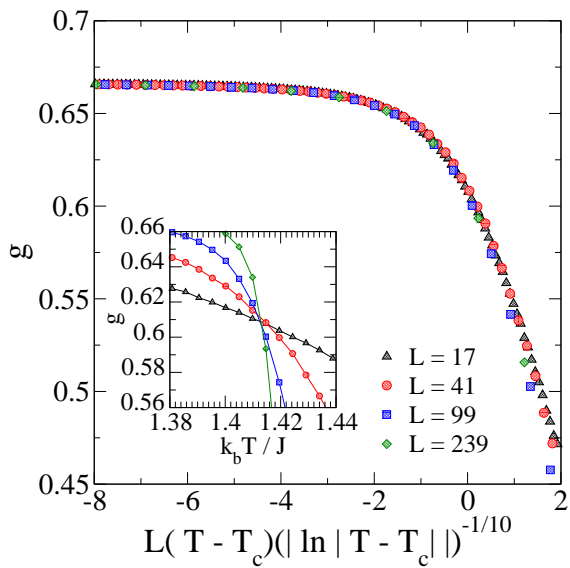}
\end{center}
\caption{(Color Online) Data collapse of the Binder Cumulant $g$
versus the scaling parameter $L^{1/\nu}(T-T_c)\left|
\ln\left|T-T_c\right| \right|^{-\hat{\lambda}}$ for different
lattice sizes $L$. Inset: Binder Cumulant versus the temperature for
different lattice sizes. The values of $L$ obey the Octonacci
sequence. We estimated the critical temperature $T_{c}\approx
1.413$, as shown in the inset, by averaging the numerical values of
the temperatures where the curves intersect each other. The best
collapse was done by using the logarithmic correction exponent
$\hat{\lambda}=1/10$. The model is in the Ising universality class
with logarithmic corrections.} \label{fig_binder_cumulant}
\end{figure}

\begin{figure}[h!]
\begin{center}
\includegraphics[scale=0.5]{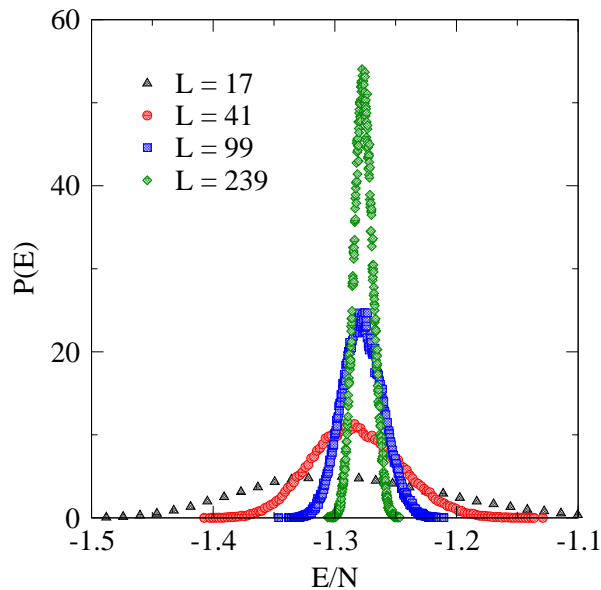}
\end{center}
\caption{(Color Online) Energy probability distribution at $T_{c}
\approx 1.41$ for different lattice sizes $L$. The values of $L$
obey the Octonacci sequence. All the curves have only one peak,
which is characteristic of a second order phase transition.}
\label{fig_energyhistogram}
\end{figure}

However, the definition of the sublattices where we can measure the
magnetization are not obvious due to the lacking of translational
symmetry. First, the lattice is not given by a superposition of
periodic sublattices, but instead, by a deterministic substitution
rule, like substitution rules in fractals (Koch's flake, Cantor
sets, Sierpinski carpets and so on). Therefore, it is not clear, in
first hand, what sublattices we should use at evaluating the
staggered magnetization. It is true that our lattice, in a given
generation, can be seen as a superposition of the lattices of the
previous generations. But, it is not clear how many sublattices one
should take, by reverting the substitution rule and considering a
number of previous generations. Also, one could argue that the
substitution rule should be reverted to the lower level where the
sublattices are the lattice spins (that is the first obvious choice)
when the staggered magnetization should read
\begin{equation}
q = \frac{1}{N} \sum_{i=1}^{N} S^0_i S_{i},
\label{pseudomagnetization}
\end{equation}
where the local phase factors are $S^{0}_{i} = \pm 1$. This can be interpreted as the system resembling, at finite temperature, its state at $T=0$. This can be seen by noting that the staggered magnetization written in Eq.(\ref{pseudomagnetization}) goes to $1$ when $T \rightarrow 0$ if $S^0_i$ is the equilibrium ground state. We have two problems when using this choice: First, i) That order parameter is clearly local, so, the order parameter should depend on the size of the lattice and ii) The ground state is degenerate, at least for Octonacci substitution rule: we do not know \textit{a priori} what ground state of the system will try to ``resemble''. We can solve i) by noting that the ground state clearly follows the substitution sequence (see caption of Fig.\ref{fig_steadystate}), so, if we have the ground state for a smaller lattice size, we can generate the ground state for the bigger lattice sizes by applying the substitution rule (so the order parameter does not depend ultimately on the system size but on the substitution rule). This even justifies why we should have to use the Pell's numbers given by ̉̉ (\ref{pellrecursionrule}), because the ground state for the infinite lattice should fit on the finite lattice. We can solve ii) by taking the averages on the size of the ordered cluster at $T=0$ in the denominator i.e. taking the averages excluding the undecided spins as already mentioned.

As we have commented in the introduction, due to quasicrystals have
no translational invariance, many anomalous properties distinct from
those of a periodic lattice are expected. Here we can see that the
long-range correlation induced by the Octonacci sequence plays an
important role in the ground state, and it has self-similar
properties (as in fractals). Also, our result is qualitatively in
accordance with some 3D quasicrystals reported in
refs.\cite{PhysRevB.57.R11047,SCHEFFER2000629}, that suggest a
possible magnetic ordering at low temperature, by measuring a
spin-glass-like transition at approximately $5.8\mathrm{K}$.
Quantitatively, instead, the quasiperiodic sequence induces
logarithmic corrections on the critical behavior in our simple
model. This can be a result of the incommensurability between
correlation lengths and the system sizes where the correlation
length of the finite system have to be taken into consideration.

\begin{figure}[h!]
\begin{center}
\includegraphics[scale=0.5]{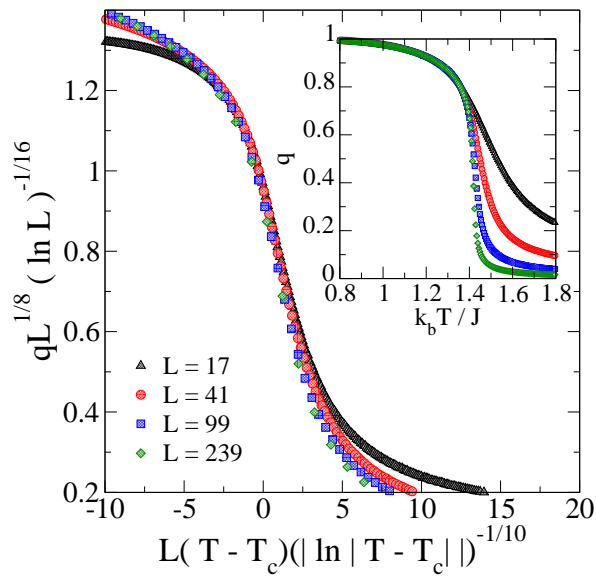}
\end{center}
\caption{(Color Online) Data collapse of the order parameter $q$,
rescaled by $L^{\beta/\nu}\left(\ln
L\right)^{\hat{\beta}+\beta\hat{\lambda}}$ versus the scaling
parameter $L^{1/\nu}(T-T_c)\left| \ln\left|T-T_c\right|
\right|^{-\hat{\lambda}}$ for different lattice sizes $L$. Inset:
order parameter $q$ as a function of temperature $T$ for different
lattice sizes $L$. The values of $L$ obey the Octonacci sequence.
The curves suggest a second order phase transition. The best
collapse is done by using the values for the logarithmic correction
exponents: $\hat{\beta} = -3/40$ and $\hat{\lambda}=1/10$. The model
is in the Ising universality class with logarithmic corrections.}
\label{fig_orderparameter}
\end{figure}

Following our previous article\cite{PhysRevE.93.042111} we estimate
the critical temperature by using the Binder cumulant $g$ given by
eq. (\ref{binder_cumulant}) in order to obtain the critical
temperature. We show the Binder cumulant in the inset of
Fig.(\ref{fig_binder_cumulant}). The critical temperature $T_{c}$ is
estimated at the point where the curves for different size lattices
intercept each other. We obtained $T_{c} \approx 1.413$. The order
of the transition is identified by the Lee-Kosterlitz
criterion\cite{PhysRevLett.65.137}, which establishes that the
thermodynamic limit of the energy probability histogram $P(E)$ can
be used to identify if a transition is continuous. The histogram,
depicted in Fig.(\ref{fig_energyhistogram}) for all lattice sizes
have only one peak, including the biggest one, and this is
characteristic of a continuous (second-order) transition.

\begin{figure}[h!]
\begin{center}
\includegraphics[scale=0.5]{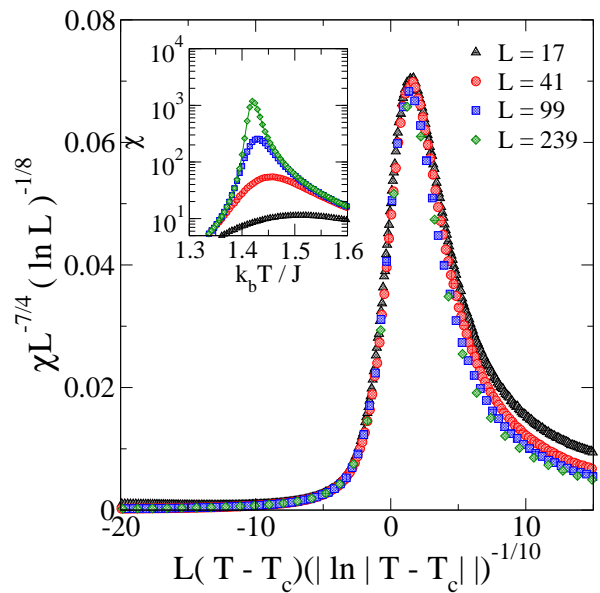}
\end{center}
\caption{(Color Online) Data collapse of the susceptibility $\chi$,
rescaled by $L^{-\gamma/\nu}\left(\ln
L\right)^{\hat{\gamma}-\gamma\hat{\lambda}}$ versus the scaling
parameter $L^{1/\nu}(T-T_c)\left| \ln\left|T-T_c\right|
\right|^{-\hat{\lambda}}$ for different lattice sizes $L$. Inset:
Susceptibility $\chi$ as a function of temperature $T$ for different
lattice sizes $L$. The values of $L$ obey the Octonacci sequence.
The susceptibility diverges at $T_{c}$ in the large lattice size
limit suggesting a second order phase transition. The best collapse
is done by using the values for the logarithmic correction
exponents: $\hat{\gamma} = 1/20$ and $\hat{\lambda}=1/10$. The model
is in the Ising universality class with logarithmic corrections.}
\label{fig_susceptibility}
\end{figure}

We display the order parameter $q$ versus temperature $T$ in the
inset of Fig.(\ref{fig_orderparameter}). The $q$ dependence suggests
the presence of a second-order phase transition in the system. We
show the data collapse by using the FSS relation written on
Eq.(\ref{q_fss}) in the Fig.(\ref{fig_orderparameter}).

\begin{figure}[h!]
\begin{center}
\includegraphics[scale=0.5]{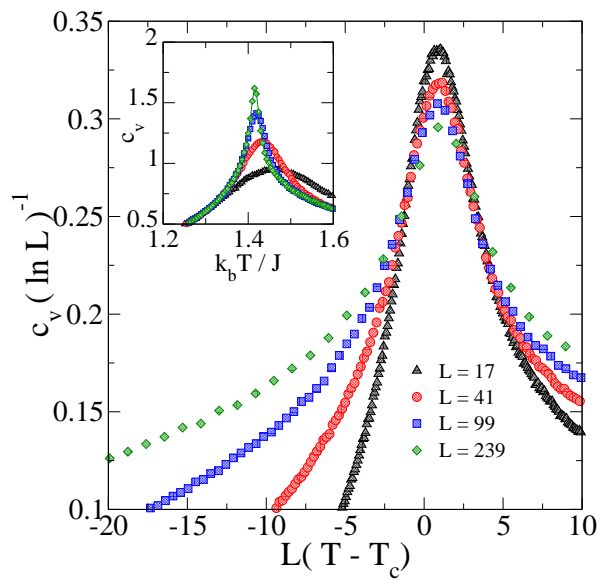}
\end{center}
\caption{(Color Online) Specific Heat $c$, rescaled by $1 / \ln L$
versus the scaling parameter $L^{1/\nu}(T-T_c)$ for different
lattice sizes $L$. Inset: Specific Heat $c$ as a function of
temperature $T$ for different lattice sizes $L$. The values of $L$
obey the Octonacci sequence. We can see that the FSS relation
without logarithmic corrections does not collapse our numerical
data.} \label{fig_specificheat-nolncorrection}
\end{figure}

\begin{figure}[h!]
\begin{center}
\includegraphics[scale=0.5]{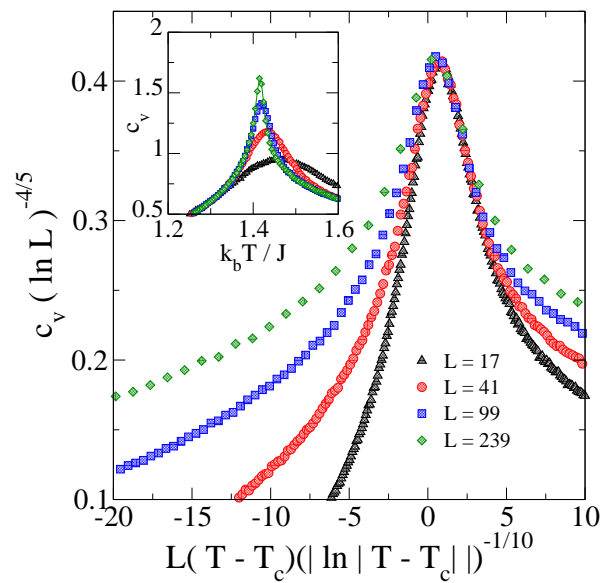}
\end{center}
\caption{(Color Online) Data collapse of the Specific Heat $c$,
rescaled by $\left( \ln L \right) ^{-\hat{\alpha}}$ versus the
scaling parameter $L^{1/\nu}(T-T_c)\left| \ln\left|T-T_c\right|
\right|^{-\hat{\lambda}}$ for different lattice sizes $L$. Inset:
Specific Heat $c$ as a function of temperature $T$ for different
lattice sizes $L$. The values of $L$ obey the Octonacci sequence.
The best collapse is done by using the values for the logarithmic
correction exponents: $\hat{\alpha} = 4/5$ and $\hat{\lambda}=1/10$.
The model is in the Ising universality class with logarithmic
corrections.} \label{fig_specificheat}
\end{figure}

Continuing the analysis of the critical exponents, we obtained the
susceptibility $\chi$ as a function of temperature $T$ in the inset
of Fig.(\ref{fig_susceptibility}). In the large lattice size limit,
the susceptibility diverges at $T_{c}\approx 1.413$. Finally, we
show the data collapse of the susceptibilities for different lattice
sizes according to FSS relation given in the
Eq.(\ref{susceptibility_fss}). All maxima are well fitted by using
the FSS relation with logarithmic corrections and the Ising critical
exponents.

Finally, we show the specific heat $c$, given by the
Eq.(\ref{specific_heat}), at the inset of the
Figs.(\ref{fig_specificheat-nolncorrection}) and
(\ref{fig_specificheat}). We estimate the $\hat{\alpha}$ exponent by
collapsing the specific heat $c$ for different lattice sizes
following the scaling relation presented in
Eq.(\ref{specificheat_fss}). We note that the maxima of the specific
heat diverges as a power of $\ln L$ as shown in
Fig.(\ref{fig_specificheat}), unlike the pure model, in which the
maxima scales as $\ln L$. Using the exponents of pure model without
logarithmic corrections does not collapse our numerical data, as
shown in Fig. (\ref{fig_specificheat-nolncorrection}). Our best
estimate for the $\hat{\alpha}$ exponent ratio is $\hat{\alpha} =
4/5$, which obeys the scaling relations for the logarithmic
correction exponents given in Eq.(\ref{fss_scaling_relations1}). As
already anticipated, this justifies \textit{a posteriori} our
choosing for the order parameter, because the critical behavior of
the $q$, its susceptibility and the specific heat preserves the
scaling relations.

\section{Conclusions}

We have presented a theoretical model with quasiperiodic long-ranged
order based on Octonacci quasiperiodic sequence, with competing
interactions, and we have obtained a critical behavior of a second
order phase transition, driven by the temperature. In the low
temperature limit we obtained an aperiodic dilute ferrite phase with
critical temperature $T_{c}\approx 1.4135$, which is different from
the model with long-ranged order based on Fibonacci
sequence\cite{PhysRevE.89.042139}. This result gives us the idea
that the aperiodic diluted ferrimagnet phase in quasicrystals
\cite{Hattori0953} could be due to the frustrated plaquettes in the
quasiperiodic order. Specifically, we have obtained the critical
exponents $\beta=1/8$, $\gamma=7/4$ and $\nu=1$ (Ising universality
class) and estimatives for logarithmic correction exponents given by
$\hat{\alpha}=4/5$, $\hat{\beta}=-3/40$, $\hat{\gamma}=1/20$ and
$\hat{\lambda}=1/10$ in the case of equal antiferromagnetic and
ferromagnetic strengths. Therefore, the quasiperiodic ordering is
marginal in the sense of introducing logarithmic corrections as in
seen for 4-state 2D Potts model.

Finally, all results presented here can be studied through artificial magnetic structures built by using nanotechnology
\cite{RevModPhys.85.1473,NATUREPHYSICS.10.670,NATUREPHYSICS.12.162}. Many artificial magnetic structures of such systems with frustrated interactions can be designed by using nanotechnology. Recent examples are systems which are known as artificial spin ices\cite{RevModPhys.85.1473}. Other exotic artificial structures, as the ``Shakti'' lattice, which displays topologically induced emergent frustration \cite{NATUREPHYSICS.10.670}, and the ``tetris'' lattice \cite{NATUREPHYSICS.12.162}, were artificially fabricated by nanotechnology lithography techniques. We hope that our work inspires experimentalists to fabricate and study this interesting system modulated by the Octonacci sequence.

\section*{Acknowledgments}

We would like to thank CNPq (Conselho Nacional de Desenvolvimento Cient\'{\i}fico e tecnol\'{o}gico) and FAPEPI (Funda\c{c}\~{a}o de Amparo a Pesquisa do Estado do Piau\'{\i}) for the financial support. We acknowledge the use of Dietrich Stauffer Computational Physics Lab - UFPI, Teresina, Brazil where the numerical simulations were performed. Also, we would like to thanks to Dr. D. H. L. Anselmo that had help us in edition and discussion of this manuscript.

\section*{References}

\bibliographystyle{iopart-num.bst}
\bibliography{textv6}

\end{document}